\documentclass[aps,prd,preprint,groupedaddress,nofootinbib,byrevtex]{revtex4-1}
\usepackage{amsmath,amssymb}
\usepackage[dvips]{graphicx}
\usepackage{mathrsfs}
\usepackage{array,booktabs}
\begin{document}
\title{
Interaction of the SN1987A Neutrino with the Galaxy   
}
\author{Kenzo Ishikawa$^{(1)}$, Terry Sloan$^{(2)}$ and Yutaka Tobita$^{(3)}$}

\affiliation{(1)Department of Physics, Faculty of Science, Hokkaido
University, Sapporo 060-0810, Japan 
}
\affiliation{(2)Department of Physics, Faculty of Science, Lancaster 
University, Lancaster , United Kingdom  
}
\affiliation{(3)Department of Physics, Faculty of Science, Hirosaki 
University, Hirosaki  , Japan 
}
\date{\today}


\begin{abstract}
 In previous publications we have shown that long-ignored approximations
 in standard model calculations could have significant implications for 
 very low mass particles such as neutrinos and photons. In particular
 we showed that, in a dilute plasma such as that in the solar corona, a
 significant decay probability of $\nu' \rightarrow \nu + \gamma$ will be
 possible as a consequence of the terms  ignored in making the
 approximations. Here the $\nu'$ and $\nu$ are high and low mass
 eigenstates of the neutrino. In this paper, we investigate the effect in
 the vicinity of an expanding supernovae remnant such  as SN1987A.  We show
 that, in the dilute plasma external to the remnant, such decays are
 possible and significant. We describe a calculation of effects of such
 decays on the anti-neutrinos from SN1987A. The calculated anti-neutrino
 energy against arrival time agrees reasonably well with that observed,
 assuming that the expansion velocity of the  remnant is $\approx 0.2 c$
 and that the plasma density is high within the expanding remnant.

\end{abstract}
\pacs{14.60.Lm, 14.60.St,14.60.Pq,13.35.Bv,14.60.Ef}
\keywords{finite-size correction, quantum interference,neutrino mass}
\preprint{
EPHOU-14-006}
\maketitle
\section{Introduction}
Anti-neutrinos from the supernovae SN1987A in the Large Magallanic Cloud(LMC) have been detected and
measured by the Baksan, IMB, and Kamiokande Collaborations \cite{Kamioka,IMB,Baksan}. The results
indicate that their mean energy falls with time. In this paper an
explanation of the phenomenon is given involving a new process.

Calculation of the transition rate, $\Gamma$, for the process   
\begin{eqnarray}
\nu' \rightarrow \nu +\gamma \label{neutrino-decay}
\end{eqnarray}
using the Standard Model (SM) of particle physics shows that it is very
small. However, the SM calculation assumes plane waves for the wave
functions of the particles which quickly become independent of the
source and the integrations are made over infinite space and momentum
\cite{Dirac,Wigner-Weisskopf,Goldberger-Watson-paper,feynmann}.
This is an approximation since  the sources of the particles involve
finite distances and sizes and the waves are not exactly plane. The difficulty of the calculation using plane waves was found sometime ago by Stueckelberg \cite{Stuckelberg} . 
Making exact calculations with the appropriate boundary conditions and
without these approximations shows that the transition probability, P,
becomes 
\begin{eqnarray}
 P=\Gamma T+ P^{(d)}
\end{eqnarray}
where T is a time interval for $ P<1$ \cite{Ishikawa-Tobita-PTEP,Ishikawa-Tobita-ANA,Ishikawa-Tajima-Tobita-PTEP}. The term $\Gamma T$ results from the SM calculation while the term $P^{(d)}$, which is time independent, is an extra term resulting from the ignored approximations in the SM calculation.   
In most SM applications involving heavy particles the probability
$P^{(d)}$ is small and undetectable. However, this is not always the case
for very low mass particles, such as neutrinos \cite{pdg} or where
$\Gamma T$ is
small
\cite{Cowsik,Pal-wolfenstein, Landau, Yang,Gell-Mann}. It
has been shown in a previous publication that if neutrinos pass through
a plasma of low enough density, such that the effective mass of the
photon is smaller than the neutrino mass difference, the probability
$P^{(d)}$ is finite and can have observable consequences. Here the photon
effective mass is $m_{\gamma,eff}=\hbar \omega_p$ where $\omega_P$ is
the plasma frequency. One such consequence is the stimulation of the
decay(equation 1) for neutrinos in the low density plasma of the solar
corona. The decay rate predicted by the theory is sufficient to account
for the heating of the corona to very high temperatures, a so-far
unexplained phenomenon  \cite{Ishikawa-Tobita-Corona}.  The effect in the solar
corona is analogous to the Hall effect for electrons in solids and it
is termed the Electroweak Hall Effect (EHE). 

The assumption of plane waves and asymptotic independence of the wave
functions in the SM calculation results in a Dirac Delta function in the
transition probability. This automatically conserves the measurable
energy for finite times. However, this is not the case for neutrinos in
a dilute plasma where $\Gamma$ is small and  $P^{(d)}$ is finite. An
interaction  energy of waves then enters through the many-body
interactions. Consequently
the total energy is conserved but the visible energy/momentum, termed
kinetic energy, are not conserved since { the interaction potential energy term} is not
accounted for \cite{Ishikawa-Tobita-Corona}.  A consequence of this is that
in the plasma which stimulates the decay in equation 1, the outgoing
$\nu'$ and $\gamma$ waves are coherent only at small angles less than of order
${m_{\nu}^2 \over E_{\nu}^2}$  where $m_{\nu}$  and $E_{\nu}$  are the
mass and energy of the decaying neutrino. 
The $P^{(d)}$ term should also exist for photons interacting via either
Thompson or Compton scattering. The photon and scattered electron waves
will be coherent and depend on the source size. Due to the large source
sizes deviations from the standard formula for these processes occur in
the far forward directions. It is shown that in the dilute plasma
external to SN1987A the coherent electron-photon wave packets move at
reduced velocity. In this way the photons from the decay of the
anti-neutrinos will be delayed in their arrival time at the Earth. This
accounts for the failure of the Solar Maximum Mission (SMM) \cite{GRS,kolb-turner,Bludman} to
detect prompt gamma rays from the decays of the anti-neutrinos from
SN1987A.
In this paper the details of the calculation of $P^{(d)}$  for the
anti-neutrinos from SN1987A are given and the spectrum of energies
and arrival times are computed and compared with the measurements.

In Section 2, the
electroweak Hall interaction and the wave functions of the decaying
neutrino are  presented. The transition amplitude and
probability are computed in Section 3, and the comparison with the
observation is made in Section 4. The summary is given in Section 5.   
\section{Electroweak Hall effect and anomalous radiative transition } 
 $P^{(d)}$ has different
 properties from  $\Gamma T$. For its computation,
we follow  the von Neumann's fundamental principle of quantum mechanics 
(FQM), that connects the probability $P$ with the state vectors, 
$P=|\langle \alpha| \beta \rangle |^2$, for normalized
 states. Since states in nature or experiments have finite sizes, the
 probability thus computed is compared with the observations. The plane
 waves, which are idealistic for 
theoretical studies, are not appropriate owing to its non-normalized 
nature.

It was found before \cite{Ishikawa-Tobita-PTEP,Ishikawa-Tobita-ANA}  that  $P^{(d)}$'s   characteristic length  is
$\lambda^{(d)}={2 \hbar c E \over m^2c^4}$, where c,m, and E are the light
velocity, mass, and the energy which depend on the absolute mass  
$m_{\nu}$ and $m_{\gamma}$ the effective photon mass $m_{\gamma}$ determined from 
the  plasma frequency. This is much longer than  the de Broglie length 
$\lambda_{dB}={\hbar \over p}$, where $\hbar ={h \over 2\pi}$ and $h$ is
the Plank constant, and $p$ is a momentum for light particles, and can be
 extremely   long. Hence, the effect appears 
 as  a macroscopic quantum phenomenon. $P^{(d)}$ is independent of
 $T $, and important in the processes for which    $\Gamma $ is very
 small  such as 
$\nu+\gamma \rightarrow \nu+\gamma, \nu +B(E) \rightarrow \nu+\gamma$ 
\cite{Landau,Yang,Gell-Mann}.  $P^{(d)}$ 
derived from the vacuum fluctuation of the  tri-angle 
electron loop which would be useful for relic neutrino observations 
\cite{Ishikawa-Tajima-Tobita-PTEP}, now is further enhanced  by  the 
electroweak Hall effect in the dilute   weak magnetized plasma.

The interaction Hamiltonian of the electroweak Hall effect is obtained
from the one-loop effect of the electrons in the magnetic field in the form,
\cite{Ishikawa-Tobita-Corona} 
\begin{eqnarray}
& & H_\text{int} =H^{Faraday}+G_{\nu,\gamma}
\int d\vec{x}\left(
\bar{\nu}(x)\left(1 - \gamma_5\right)\gamma_\mu \nu'(x)\right) 
\epsilon^{\mu\alpha\beta} {\partial_{\alpha} } A_{\beta},\\
& &G_{\nu,\gamma}=\frac{G_F}{\sqrt{2}}e{{\nu^{(4)} \over 2\pi}},\nu^{(4)}= {2\pi \hbar n_e \over eB} \nonumber
\end{eqnarray}
where $H^{Faraday}$ is the Chern-Simons term of the electromagnetic
potential, $G_F$ is  Fermi
coupling constant, $n_e$ is the electron density, and $B$ is the
magnetic field. Electrons  in each Landau level give independent
contributions in the loop integral and  the coupling strength  is
proportional to  the filling
fraction and 
sizable in a dilute plasma in a weak magnetic field.  $H_{int}$ 
derives from the quantum fluctuation of the electrons in the magnetic
field, and $H^{Faraday}$ gives the Hall effect for the electromagnetic
current and the Farady rotation of the electromagnetic waves, which is
useful for the measurement of the magnetic field in the Galaxy \cite{galaxy-magnetic-field}.  The
rest gives the neutrino-photon  interaction, and is applied to the
neutrinos from SN1987A.  For  the
magnetic field in the $z=x_3$-direction, $(\mu,\alpha,\beta)$ is the space
perpendicular to $x_3$-axis, i.e.,  $(0,1,2)$.

 The 
 Schr\"{o}dinger equation  for the processes Eqs. $(\ref{neutrino-decay})$ is
\begin{align}
 i\hbar\frac{\partial}{\partial t}|\Psi(t)\rangle &= H|\Psi(t)\rangle=\left(H_0 + H_\text{int}\right)|\Psi(t)\rangle,\label{shrodinger-eq}\\
H_0 &= \int d\vec{x}\sum_{l=e,\mu} \left( \bar{l}(x)\left(\vec{\alpha}\cdot\vec{\nabla} + \beta m_l\right)l(x) + \bar{\nu}_l
(x)\left(
\vec{\alpha}\cdot\vec{\nabla} + \beta m_{\nu_l}\right)\nu_l(x)\right).\nonumber
\end{align}
  The case of no-mixing is considered first. Normalized 
 $|\Psi(t)\rangle$ evolving from one neutrino state of momentum ${\vec
 p}_{\nu}$ coming from a source  size of $\sigma_{\nu}$ at $t=0$ is
\begin{align}
& |\Psi(t),\vec{p}_{\nu}\,\vec X_{\nu},\sigma_{\nu}\rangle =
 a_0(t)|\vec{p}_\nu ,\vec X_{\nu},\sigma_{\nu}
\rangle+\int
 d\vec{p}_e d\vec{p}_{\gamma} a_1(t,\vec{p}_{\nu'},\vec
 X_{\nu'},\sigma_{\nu'},\vec{p}_{\gamma})|\vec{p}_{\nu'},\vec
 X_{\nu'},\sigma_{\nu'},\vec{p}_{\gamma},\sigma_{\gamma}\rangle+O(G_F^2),\nonumber \\
& a_0(t) = {(1+\zeta(t))}^{-1/2}e^{-i\frac{E_\nu}{\hbar}t },\\ 
&
 a_1(t,\vec{p}_{\nu'},\vec X_{\nu'},\sigma_{\nu'},\vec{p}_{\gamma},\sigma_{\gamma}) = (1+\zeta(t))^{-1/2} e^{-i\frac{E_\nu}{\hbar}t}\frac{e^{-i\frac{\omega}{\hbar}t} 
- 1 }{\omega  }\langle \vec{p}_{\nu'},\vec X_{\nu'},\sigma_{\nu'}
\vec{p}_{\gamma},\sigma_{\gamma}|H_\text{int}|\vec{p}_\nu,\sigma_{\nu} \rangle, \label{weight}
  \\
& \zeta(0)=1, \zeta(t) =\zeta <1;(t >0) \nonumber
\end{align}
in the lowest order of $G_F$, where $\omega =  E_{\nu'} +
E_{\gamma} - E_\nu$, and  
\begin{eqnarray}
& &\zeta(t)= \int_{\omega \neq 0}  
 d\vec{p}_{\nu'}d\vec{p}_{\gamma}  
({2\sin{\omega t \over 2\hbar} \over \omega})^2| \langle \vec{p}_{\nu'},\sigma_{\nu},
\vec{p}_{\gamma},\sigma_{\gamma}|H_\text{int}|\vec{p}_\nu,\sigma_{\nu} \rangle|^2. \label{integration}
\end{eqnarray} 
$\sigma_{\gamma}$ is the range in space covered by  the wave
function that the  photon interacts with in the microscopic process, which
is called the photon wave packet hereafter. That is large and 
the waves deviate slightly from the plane waves.   
$\sigma_l$ stands for the wave packet size
for particle l. $\sigma_{\nu}$ for the initial neutrino is
determined by the size of the star, and  $\sigma_{\nu'}$ and
$\sigma_{\gamma}$ for the final neutrino or photon is
determined by the range in space covered by  the wave function of microscopic object which
these interact. $\sigma_{a+b}$ stands for the cross section of the
process $a+b$. In most of this paper $\sigma_{\nu}$, $\sigma_{\nu'}$, and
$\sigma_{\gamma}$ are considered large.   $\Gamma$ is proportional to
$m_{\nu}^5$ and practically $\Gamma$ is negligibly small,
$\tau = \infty$, hence $a_0(t)$ and
$a_1(t)$ have no exponential damping factor in time of the
Weisskopf-Wigner formula \cite{Wigner-Weisskopf}. 
$\zeta(t)$  converges due to the wave packets and approaches  
a constant $\zeta$ rapidly.
 The
square of norm of $\nu$ and $\nu'+\gamma$ at $t \rightarrow \infty$   is
$(1+\zeta)^{-1}$ and $(1+\zeta)^{-1} \zeta$, where the former is the 
survival probability for the parent.   
$P^{(d)}$ is independent of $\Gamma$, and $\Gamma=0$ but $P^{(d)} \neq 0$.
\cite{Ishikawa-Tobita-PTEP,Ishikawa-Tobita-ANA,Ishikawa-Tajima-Tobita-PTEP}. 

The integral Eq.$
(\ref{integration})$ from the region $\omega \neq 0$ diverges, 
if all the states are plane waves,  and has been 
considered not to be relevant to a physical
phenomenon. Accordingly any physical
quantity has not been derived from the region $\omega \neq 0$.  However
the divergence is inherent to the plane waves and disappears in realistic
situations. The final state 
is expressed by a wave packet of finite size in the physical 
process, where the decay product interacts with other microscopic states
of the finite  range in space.   The integral is convergent  then and  the probability 
derived from the region $p_{\nu'} \rightarrow \infty$ and  that from the
  region $p_{\gamma} \rightarrow \infty$ become finite and possesses  
  universal properties.  They determine the probability of 
the events  that the photon or the
  neutrino is measured or that they make reactions. In both cases, the 
unmeasured state, i.e., the $\nu'$  in the   
former and the $\gamma$ in the latter includes  the state  $p 
  \rightarrow \infty$ inherent in a relativistic invariance.  Because the waves in two regions are different,
  they are independent each other, and are computed in the next section.  
\section{The transition amplitude  }
The probability amplitude of  
$\nu \rightarrow \nu' +\gamma$ is determined by the
initial and the final wave functions at finite time interval
following FQM using $S[T]$, the matrix 
element which is determined from Eq.$(\ref{weight})$. 
  $S[T]$ satisfies $[S[T],H_0] \neq 0$ due to  the overlap of waves, 
whereas the standard  S-matrix, $S[\infty]$, satisfies $[S[\infty],H_0]
= 0$ from   the asymptotic boundary condition at $T\to \infty$
\cite{Goldberger-Watson-paper}. $S[\infty]$  is
useful for computing $\Gamma $ but useless for $P^{(d)}$.
 $S[T]$  is  formulated with M{\o}ller operator  \cite{Ishikawa-Tobita-PTEP,
  Ishikawa-Tobita-ANA}, and the normalizable wave
 functions,   wave packets 
that are localized in space  \cite
 {LSZ} and 
specified by   their centers in the momentum and coordinate. 
\cite{Ishikawa-Shimomura,Ishikawa-tobita-ptp}.  
The amplitude for  an initial neutrino    
denoted as  $|\nu\rangle$ to   final neutrino $\nu'$ and a photon
of the momentum ${\vec p}_{\gamma}$,   
$ |\nu\rangle = |\vec{p}_\nu,\vec{X}_\nu,T_\nu;\sigma_{\nu}\rangle,
\ |\nu' \gamma \rangle = |\vec{p}_{\nu'},\vec X_{\nu'},\sigma_{\nu'};
\vec{p}_{\gamma},T_{\gamma}\rangle;  \delta p = p_\nu-p_{\nu'}-p_{\gamma} $
 is
\begin{align}
& \mathcal{M}=G_{\nu,\gamma}
 \varrho_\nu\varrho_{\nu'}\varrho_{\gamma} f I(\delta p),f=\bar{u}(\vec{p}_{\nu})\gamma^\rho(1-\gamma_5)u(\vec{p}_{\nu'})
\epsilon^{mag}_{\rho}(p_{\gamma}) ,\label{amp-3}\\
&  \varrho_\alpha =
 \left(\frac{m_\alpha}{(2\pi)^3E_\alpha}\right)^\frac{1}{2}(\alpha=\nu,\nu'),\ \varrho_\gamma =
 \left(\frac{1}{(2\pi)^3 2E_\gamma}\right)^\frac{1}{2}, \nonumber
\end{align}
\begin{eqnarray}
\epsilon^{mag}_{\rho}(p_{\gamma}) =\langle matter' |\epsilon^{\rho\alpha\beta}\partial_{\alpha}A_{\beta}|matter \rangle,
\end{eqnarray}
where $\bar u(p_{\nu})$, $u(p_{\nu'})$, and $\epsilon^{mag}_{\rho}(p_{\gamma})$ are the
spinors of the neutrinos and the photon coupling vector with matter in the magnetic field, and
$(\rho,\alpha,\beta)$ is the three dimensional space of Eq.$(\ref{shrodinger-eq})$.
In Eq.$(\ref{amp-3})$ the last term  is,  
\begin{align}
& I(\delta p)=\int_{T_\nu}^{T_{\nu'}} dt\int d\vec{x} e^{i \phi_{\gamma}(x)}
w(x,X_\nu;\sigma_\nu) w^{*}(x,X_{\nu'};\sigma_{\nu'}) \\
&\phi_{\gamma}(x,\vec{p}_\gamma)= E(\vec{p}_\gamma) t -
 \vec{p}_\gamma\cdot\vec{x}, \nonumber
\end{align}
where the wave function is
\begin{eqnarray}
& &\omega(x,X_{\alpha},\sigma_{\alpha})=({4\pi \over \sigma_{\alpha}})^\frac{3}{4}e^{-{1 \over 2\sigma_{\alpha}}({\vec x}-{\vec
 X}_{\alpha}-{\vec v}_{\alpha}(t-T_{\alpha}))^2 -i\phi_{\alpha}(x,{\vec
 p}_{\alpha})} \\
& &\phi_\alpha(x,\vec{k}_\alpha)
= E(\vec{k}_\alpha)(t-T_\alpha) - \vec{k}_\alpha\cdot(\vec{x}-\vec{X}_\alpha),\ (\alpha=\nu,{\nu'})\nonumber.
\end{eqnarray}
The photon's coupling with matter through the magnetic coupling expressed by a normalized polarization vector $\epsilon_{\rho}^{mag}(p_{\gamma}) $ and a coupling strength
$f_{\gamma}$
\begin{eqnarray} 
\epsilon_{\rho}^{\alpha\beta}  \partial_{\alpha}A_{\beta}(p_{\gamma})=h_{\gamma}{\epsilon_{mag}}^{\rho}(p_{\gamma}), h_{\gamma}=\sqrt {\frac{2 p_{\gamma}^2}{3}},
\end{eqnarray}
of satisfying 
\begin{eqnarray}
\sum_{\rho} |\epsilon_{\rho}^{\alpha\beta}(p_{\gamma})_{\alpha}\epsilon_{\beta}(p_{\gamma})|^2=|\sum_{\rho}|h_{\gamma}{\epsilon_{mag}}^{\rho}(p_{\gamma})|^2, 
\end{eqnarray}
where $\epsilon_{\beta}(p_{\gamma})$ is the photon's polarization vector. The spreading of wave packet at  large $|t-T_{\nu_e}|$ now is negligible
\cite{Ishikawa-Tobita-ANA}.

The amplitude Eq.$(\ref{amp-3})$ is almost identical to that
of the plane waves, but now the wave function is  normalizable and the time interval is finite $T$.  The overlapping waves  interact each others, and  the S-matrix $S[T]$ satisfies  $[S[T],H_0] \neq 0$.   Thus the transition  to the kinetic energy non-conserving states is included  and 
 its  probability $P^{(d)}$ is computed    following the FQM.  

\subsection{Transition probability}
The vector index in Eq.$(\ref{amp-3})$ is in  $(0,1,2)$, and  the
the spin 
average $
\sum_{\text spin}|f|^2=\frac{2}{3}{2^4 \over 4 m_{\nu} m_{\nu'}} (\tilde p_{\gamma})^2 (\tilde p_{\nu'} \cdot
 \tilde p_{\nu}-\frac{3}{2} p_{\nu_i}\cdot p_{\nu_j})$, where $\tilde p$ is a vector in this  three
 dimension and  a  scalar products is that of the same three dimensional
 subspace.  The probability for the event that the
 neutrino is measured 
 or interacts,  $P^{(d)}_{\nu}$,  and
that for the photon, $P^{(d)}_{\gamma}$,   are
computed following the method
\cite{Ishikawa-Tobita-PTEP,Ishikawa-Tobita-ANA, Ishikawa-Tajima-Tobita-PTEP}, 
\begin{eqnarray}
P= \int  d{\vec p}_{\nu'} \frac{d{\vec X}_{\nu'}}{(2\pi)^3} d{\vec p}_{\gamma} |\mathcal M|^2  .
\end{eqnarray}
\subsubsection{ Interaction of neutrino and photon waves with a large wave of matter }

{\bf Neutrino probability $P^{(d)}_{\nu}$ }

A nucleus or atom in galaxy has a large mean free path due to the low
density, and is expressed by the  range in space
covered by    $\sigma_{\nu'}$, which is different from that of the initial
neutrino $\sigma_{\nu}$ determined by the size of the star and   satisfies  $\sigma_\nu\ll\sigma_{\nu'}$. The probability is written
with the smallest wave packet, $\sigma_{\nu}$ now. The photon is not measured 
and integrated over the positive energy region. Hence  $P^{(d)}_{\nu}$
is computed with the correlation function, 
\begin{align}
\Delta_{\nu,\gamma}(\delta x)=\frac{2}{3} \frac{1}{(2\pi)^3}\int
 \frac{ d \vec{p}_{\gamma}}{2E_{\gamma}}
2^4 (\tilde p_\gamma)^2 
(\tilde p_{\nu} \cdot \tilde p_{\nu'}-3/2 p_{\nu_i} \cdot p_{\nu_j}) e^{-i(p_{\nu}-p_\gamma)\cdot \delta x}. \label{correlation-1}
\end{align}
The light-cone singular term inherent to the
 relativistic system from states $\omega = \infty$ in
 Eq.$(\ref{integration})$  couples with $\Delta_{\nu,\gamma}(\delta x) $
 and gives the leading contribution, 
\begin{align}
 &P^{(d)}_{\nu}  = \frac{2}{3}{1 \over (2\pi)^3}G_{\nu,\gamma}^2 
{1 \over
 E_{\nu}}2^4 \sigma_{\nu} \int\frac{d\vec{p}_{\nu'}}{E_{\nu'}}
(\tilde p_\nu-\tilde p_{\nu'})^2  (\tilde p_{\nu} \cdot \tilde
 p_{\nu'}-p_{\nu}\cdot p_{\nu'}) T\tilde
 g(\omega_{\nu}T),\label{diff-rate1}
\end{align}
where 
 $\omega_{\nu}=\frac{m_{\nu}^2}{2E_{\nu}}$, and  $v_{\nu} = c=1 $ and the electron
mass was  neglected, and the fact that  the wave packet vanishes at
$(t-T_{\nu})^2-({\vec x}-{\vec X}_{\nu})\leq 0$
\cite{Ishikawa-Shimomura} is not important now, and is
 ignored. The asymptotic behavior of $\tilde g (\omega_{\nu}T)$ given in
 Appendix is substituted.

The phase space is in the region
\cite{Ishikawa-Tobita-ANA}, $
 2p_\nu\cdot p_{\nu'}\leq m_{\nu'}^2+m_{\nu'}^2-m_{\gamma}^2$
and the integral over the angle $\theta$ between the
momenta of $\nu$ and $\nu'$ is made in the region,
\begin{align}
1-\cos \theta \leq \frac{1}{2 E_{\nu}
 E_{\nu'}}[(1-\frac{E_{\nu'}}{E_\nu})m_{\nu}^2+
 (1-\frac{E_\nu}{E_{\nu'}})m_{\nu'}^2 - \frac{m_\nu^2m_{\nu'}^2}{2 {E}_\nu
 pE_{\nu'}}-m_{\gamma}^2]. \label{phase-space1}
\end{align}
We have the total probability expressed by  the integral over the momentum
fraction $x=\frac{|{\vec p}_{\nu'}|}{ |{\vec p}_{\nu}|}$ 
\begin{align}
 &P^{(d)}_{\nu}  = {1 \over (2\pi)^3} G_{\nu\gamma }^2 \sigma_{\nu}
 E_{\nu}^4 F_{\nu}(\xi),\xi=(m_{\nu}/m_{\nu'})^2,\label{diff-rate2} \\
& F_{\nu}(\xi)=2^6 \int_{1/\xi}^1 dx x(1-x)^3(x \xi-1)\rightarrow {8 \over 15}\xi  ;\xi \rightarrow \infty.
 \nonumber 
\end{align}
$P^{(d)}_{\nu}$ has unique properties; that is proportional to  the
range in space covered by  the initial neutrino $\sigma_{\nu}$, the fourth power of the neutrino energy $E_{\nu}^4$, and 
the neutrino mass-squared ratio $\xi$. Here $\sigma_{\nu}$ is $\pi \times
R^2$, where $R$ is the radius of the exploding star, and is  a large
macroscopic value.  The 
average fraction, $\langle x
\rangle =3/7$ at $ \xi \rightarrow \infty $ of about $0.5$ is due to 
the modified phase space 
Eq.$( \ref{phase-space1})$ that includes the
region satisfying the inequality. The absolute value of momentum can deviate
from the initial value despite 
$\theta \approx 0$. It is noted that  a naive value of $\Gamma$ for a
weak process $G_F^2 m_{\nu}^5$ is negligible, but $P_{\nu}^{(d)}$ is
different and can give significant effects. 
The $E_{\nu}^4$ is larger than $m_{\nu}^4$ by the factor  
$({E_{\nu}
\over m_{\nu}})^4$, which becomes now $({10^7 \over 10^{-1}})^4
=10^{32}$.  The enhancement due to the electroweak
Hall effect is further amplified by the Lorentz non-invariance. 

{\bf Photon  probability $P^{(d)}_{\gamma}$  }

The probability that the photon interacts with matter is expressed by
their wave functions, and the range in space covered by  them  
is determined by that mean free path.
They satisfy   $\sigma_{\gamma} \approx \sigma_{\nu'} \gg \sigma_{\nu}$.
The neutrino momentum in 
the final state is integrated in the  phase 
space  \cite{Ishikawa-Tobita-ANA} is replaced with,
$ 2p_\nu\cdot p_{\gamma}\leq m_\nu^2-m_{\nu'}^2+m_{\gamma}^2$.
Now $\tilde \omega_{\gamma}=E_{\gamma}(1-\cos \theta)+{m_{\gamma}^2
 \over 2E_{\nu}}$, where $\theta$
 is the angle between ${\vec p}_{\nu}$ and ${\vec p}_{\gamma}$, which  
is almost zero from Eq.$(\ref{angle})$ discussed later.  It follows that
for a large  $T$,
\begin{align}
 &P^{(d)}_{\gamma}  = {1 \over (2\pi)^3} G_{\nu\gamma }^2  {1 \over
 E_{\nu}}2^4 \sigma_{\nu} \int\frac{d\vec{p}_{\gamma}}{E_{\gamma}} 
(\tilde p_\nu-\tilde p_{\gamma})\cdot \tilde p_{\gamma}  \tilde p_{\gamma}\cdot \tilde p_{\nu}( T\tilde
 g(\tilde \omega_{\gamma} T)).\label{diff-rate4}
\end{align}
Substituting the asymptotic form of $\tilde g(\tilde \omega_{\gamma}T)$,
and integrating  over the the region 
\begin{align}
 1-\cos \theta \leq \frac{1}{2 p_{\nu}
 p_{\gamma}}[(1-\frac{p_\gamma}{p_\nu})m_{\nu}^2-
 m_{\nu'}^2 +m_{\gamma}^2], \label{angle}
\end{align}
we have the total probability expressed by  the integral over the momentum
fraction $x={p_{\gamma} \over p_{\nu}}$ 
\begin{align}
 &P^{(d)}_{\gamma}  =  {1 \over (2\pi)^3}  G_{\nu\gamma }^2 \sigma_{\nu} 
 E_{\nu}^4  F_{\gamma}(\xi) \text {log} \frac{2 E_{\nu}^2}{ m_{\gamma}^2},\label{diff-rate6} \\
&  F_{\gamma}(\xi)=2^4 \int_{1/\xi}^{1-1/\xi  } dx x(1-x)(x-
 1/\xi) \rightarrow {8 \over 3}  ;\xi \rightarrow \infty \nonumber, 
 \end{align}
which is proportional to the range in space covered by  the initial neutrino 
$\sigma_{\nu}$ and the log factor of the initial momentum over the mass,
the fourth power 
of the neutrino energy $E_{\nu}^4$. The probability is enhanced  over 
the normal case by a factor  
$({E_{\nu}
\over m_{\nu}})^4$ and by the large log factor of the momentum $\text
{log} \frac{2 E_{\nu}^2}{ m_{\gamma}^2} \approx 10^2$. For the
small $T$, we have 
\begin{align}
 &P^{(d)}_{\gamma}  = {1 \over (2\pi)^3}  G_{\nu\gamma }^2 {1 \over
 E_{\nu}}2^4  \int\frac{d\vec{p}_{\gamma}}{E_{\gamma}} ( {4 \over
 p_{\gamma} {\tilde \omega_{\gamma}}^2}) 
(\tilde p_\nu-\tilde p_{\gamma})\cdot \tilde p_{\gamma}  \tilde p_{\gamma}\cdot \tilde p_{\nu},\label{diff-rate5}
\end{align}
and 
\begin{align}
 &P^{(d)}_{\gamma}  =  {1 \over (2\pi)^3} G_{\nu\gamma }^2  {1 \over m_{\gamma}^2}  
 E_{\nu}^4  F_{\gamma}(\xi) ,\label{diff-rate8} \\
&  F_{\gamma}(\xi)=2^4 \int_{1/\xi}^1 dx x(1-x)(x-
 1/\xi) \rightarrow {8 \over 3}  ;\xi \rightarrow \infty \nonumber, 
 \end{align}
which is independent of $\sigma_{\nu}$. $m_{\gamma}$ is extremely small
and ${1 \over m_{\gamma}^2 }\approx (2 \times 10^8)^2 {\text m}^2 $
for $m_{\gamma} =10^{-15} \text {eV}$.
Thus $P^{(d)}$ of Eq.$(\ref{diff-rate8} )$ is not very different from
that of  Eq.$(\ref{diff-rate6} )$.
The average
energy fraction of the final neutrino, $\langle x
\rangle =1/2$ at $ \xi \rightarrow \infty $  from the same reason as the
previous case. 

{\bf Summary of $P^{(d)}$}

The overlapping waves of the initial and final neutrinos result
to  $P^{(d)}_{\nu}$ from the kinematic 
region $p_{\nu'} \leq p_{\nu}; p_{\gamma} \rightarrow \infty$,   and
those of  the initial neutrino and the final photon  result to  
$ P^{(d)}_{\gamma}$ from  the region $p_{\gamma} \leq p_{\nu},~p_{\nu'}
\rightarrow \infty$. They are from different kinematic regions, and  are added. 

The satellite galaxy LMC is likely to have magnetic fields and electron densities similar to our own galaxy, the Milky Way. These will also affect the neutrino by the electroweak Hall effect so that the total value of $P^{(d)}$ for the neutrinos will be the sum of the conventional probabilities in the LMC and the Milky Way Galaxy. We denote $c_1$ for the Galaxy and $c_2$ for LMC,
 and write the probability respectively as 
\begin{align}
 &P^{(d)}(c_i)=P^{(d)}_{\nu}(c_i)+P^{(d)}_{\gamma}(c_i)  = {1 \over (2\pi)^3} {e^2 G_F^2 \over 2}({\nu^{(4)}(c_i) \over
 2\pi})^2 \sigma_{\nu} 
 E_{\nu}^4 F \label{diff-rate7} \\
&F= F_{\gamma}(\xi) \text {log} {2 E_{\nu}^2 \over
 m_{\gamma}^2}+F_{\nu}(\xi) . \nonumber  
 \end{align}
$F$ is around $ F \approx  10^3$ in the Galaxy 
, where   $m_{\gamma}c^2 = 10^{-16}{\text eV}$, 
   $p_{\nu} = 20{\text
{MeV}}$, and $\xi=10^{3}$ are used.   In LMC, the density and the magnetic field are not known well
and are left as parameters.  
The $P^{(d)}$ is the
sum of those of the Galaxy and LMC,
\begin{eqnarray}
P^{(d)}=P^{(d)}(c_1)+P^{(d)}(c_2)= {1 \over (2\pi)^3 }
{e^2  G_F^2 \over 2}\sigma_{\nu} 
 E_{\nu}^4 F \sum_i(({\nu^{(4)}(c_i) \over
 2\pi})^2)  \label{diff-rate9} 
\end{eqnarray}
Here $\sigma_{\nu}$ depends on the radius. That  varies slowly with the
radius and the average value appears in the final expression. There is
no contribution to $P^{(d)}$ from the region $E_{\nu'} \rightarrow \infty, E_{\gamma} \rightarrow \infty $.

\subsubsection{Survival probability}

The initial neutrino  lowers the flux due to the transition $\nu
\rightarrow \nu'+\gamma$.  
From the unitarity  
\begin{eqnarray}
\langle \nu|S[\text T]^{\dagger} \nu \rangle \langle \nu|S[\text T]|\nu \rangle +
\langle \nu| S[\text T]^{\dagger}| \nu',\gamma\rangle \langle \nu',
\gamma|S[\text T]| \nu \rangle =1,
\end{eqnarray}
where the second term in the left-hand side is computed from $P^{(d)}$,
the probability that the initial neutrino remains is given by   
 \begin{eqnarray}
|\langle \nu|S[\text T] \nu \rangle|^2 =
1-|\langle  \nu',\gamma | S[\text T]| \nu  \rangle|^2. 
\end{eqnarray}
For $ P^{(d)} \ll 1$, the correction to the norm of the initial
state is negligible and 
\begin{eqnarray}
|\langle \nu',\gamma | S[\text T]| \nu \rangle|^2= P^{(d)}.
\end{eqnarray}
For a larger $ P^{(d)}$, including the norm's correction,
we have   
\begin{eqnarray}
|\langle  \nu',\gamma| S[\text T]| \nu  \rangle|^2= {P^{(d)}
 \over 1+P^{(d)}}.
\end{eqnarray}
The survivale probability of the initial neutrino and the
probability of the produced photon  
\begin{eqnarray}
& &P(\nu \rightarrow \nu)= {1
 \over 1+P^{(d)}}, \label{survivable-probability}\\
& &P(\nu \rightarrow \gamma)= {P^{(d)}_{\gamma}
 \over 1+P^{(d)}} \nonumber
\end{eqnarray}
will be compared with the observations.

\subsubsection{\label{section:diffraction-mixing}Mixing effect}

  There are three neutrinos and they mix each others. For 
mass eigenstates $\nu_i(x);i=1,3$ of the masses  $m_{\nu_i}$, and the
mixing matrix $U_{\alpha,i}$, the flavor neutrino fields $\nu_l(x)$ 
in Eq. \eqref{shrodinger-eq} are the linear combination 
\begin{align}
\nu_l(x) = \sum_{i}U_{l,i}\nu_i(x),\ l = e,\ \mu, \tau,
\end{align}
where the best-fit values of mixing angles given in Ref. \cite{pdg}
\begin{align}
& \sin^22\theta_{12}=0.846\pm0.021,\nonumber \\ 
&\sin^22\theta_{23} = 0.999^{+0.001}_{-0.0018}\text{ (normal hierarchy)},\ 
\sin^22\theta_{23} = 1.0000^{+0.000}_{-0.017}\text{ (inverted)},
 \nonumber \\
&\sin^22\theta_{13} =(9.3\pm 0.8)\times 10^{-2}, \nonumber
\end{align}
are used and CP violation phase $\delta_{CP}= 0$ is assumed.
The amplitude that the mass eigenstate $i$ makes the radiative transition is   
\begin{align}
 \mathcal{M}_{i,\nu_e} = \mathcal{M}(\nu,i)U_{\nu_e,i}^*,
\end{align} 
where the neutrino species is not specified in the final state.
Thus the probability that the electron neutrino decays to a neutrino 
and a photon through  $P^{(d)}_{\gamma}$
is given by the factorized form,
\begin{align}
{P}_{e}^{(d)} 
= P^{(d)}_{\gamma,i} |U_{e,i}|^2. 
\end{align}
The mixing modifies the probabilities slightly. We use a correction
factor $1/2$. 

\subsubsection { Reactions of the decay products with a nucleus or an atom in the detectors}
The probability that the the neutrino or the photon directly 
detected with the detector on the earth is proportional to
the range of space covered by the bound states.   
 The wave functions of the nucleus or atoms are of microscopic sizes, and  
 $P^{(d)}$ proportional to these small sizes are much smaller 
than the previous cases.

 For the event that the neutrino is detected,  $\sigma_{\nu'}$ is  that
 of 
nucleus of the
order $\pi \times 10^{-30} \text{meters}^2$, and  
$\sigma_{\nu'} \ll \sigma_{\nu}$. Accordingly the probability is    smaller than
Eq.$(\ref{diff-rate2})$ by the ratio $\sigma_{\nu}^{nucl} /\sigma_{\nu} \leq 10^{-50}$.
Similarly for the event that the decay photon directly reacts with an atom in 
the  gamma ray detector, which is a bound atom, $\sigma_{\gamma}^{atom}$ is the atomic size of the
order $\pi \times 10^{-20} \text{meters}^2$, and $\sigma_{\gamma} \ll \sigma_{\nu}$. The probability is  
  smaller than
Eq.$(\ref{diff-rate5})$ by the ratio $\sigma_{\gamma}^{atom}
/\sigma_{\gamma} \leq 10^{-40}$. This is also negligibly small. 

During the
long travel, actually, the photon interacts with the electrons with the 
strength of Quantum Electrodynamics(QED), and  is affected 
by these reactions in the Galaxy. This effect is studied later.

\section{Comparison with the SN1987A neutrino}
We study the events of antineutrino from SN1987A observed in the ground
detectors. The number of the events is proportional to the survival
probability $P(\nu \rightarrow \nu)$, which depends upon the size
$\sigma_{\nu}$. $\sigma_{\nu}$ shows the size of the area that the
neutrino bursts takes places, which may be the size of the core of 
Supernovae, or the size of shock wave front. A current understanding
based on the numerical simulations shows the latter of the  
velocity of about one tenth of the light velocity is favored.

\subsection{Detection of the neutrino and the prompt gamma}
We introduce  the radius $R$ of the relation  
\begin{eqnarray}
\sigma_{\nu}= \pi R^2,
\end{eqnarray}
and express hereafter the probability with it. This $R$ may be around 
$10^{4}$ meters for  the supernove core or $10^{7}-10^{8}$  meters for the 
expanding shock wave.

The probability in  Eq.$(\ref{diff-rate8})$ is written as, 
\begin{eqnarray}
& &P^{(d)}_{\gamma}(R)=({R \over R_0})^2, \\
& &R_0=1.37 \times 10^{10}  ({F \over 10^3} )^{-1/2} ({20MeV \over
 E_{\nu}})^{3/2} r\\
& & r={0.4 \over  \sqrt{\nu^{(4)}(G)^2+\nu^{(4)}(LMC)^2}},
\end{eqnarray} 
where the units ${1 \over \text{meter}^3}$ and Tesla are used for $n_e$
and  $B$. For $\nu^{(4)}(LMC)=0$, $r=0.4$ corresponds to  $B(G)=10^{-10} \text{Tesla}$ and $n_e(G)=10^4
\text{meter}^{-3}$. Thus the survival probability at the earth is  
\begin{eqnarray}
P( \nu \rightarrow \nu)= {1 \over 1+({R \over R_0})^2}, R \geq R_0.
 \label{survivable }
\end{eqnarray} 

The probability of the neutrino reaction  with the nucleus in the
detector is determined by the standard cross section. Hence  using
the flux of the neutrino at the SN1987A,  $\phi(E_{\nu};SN1987A)$, 
the probability of the event that the  SN1987A neutrino is detected at
$t$, $N_{\nu}(t)$, is written as
\begin{eqnarray}
& &N_{\nu}(E_{\nu},t)= N_{\nu}^{(0)}\sigma_{\nu+nucleus} n_{nucleus}
  L(\nu),
 \label{neutrino-detection} \\
& &N_{\nu}^{(0)}=\phi(E_{\nu};SN1987A)  P( \nu \rightarrow \nu),   \nonumber
\end{eqnarray}
where $\sigma_{\nu+nucleus}$ is the neutrino nucleus cross section, 
$n_{nucleus}$ is the nucleus density, and $ L(\nu)$ is the total volume
of the detector.  The flux is modified from the naive value
$\phi(E_{\nu};1987A)$ to $N_{\nu}^{(0)}$ by $  P( \nu \rightarrow \nu)$ in the
Galaxy, and  will be compared with the observations. 

The   probability for the prompt gamma  to be detected simultaneously
with the neutrino  is
\begin{eqnarray}
& &N_{\gamma}(E_{\gamma},t)=N_{\gamma}^{(0)}  
\sigma_{\gamma+nucleus} n_{nucleus}  L(\gamma), \label{gamma-detection} \\
& &N_{\gamma}^{(0)}=\phi(E_{\nu};SN1987A) P( \nu \rightarrow
\gamma),  \nonumber
\end{eqnarray}
where $\sigma_{\gamma+nucleus }$ is the
gamma nucleus cross section, $ L(\gamma)$ is the total volume  of the detector, and others are the same as
Eq.$(\ref{neutrino-detection})$. 
The ratio $\sigma_{\gamma+nucleus}/ \sigma_{\nu+nucleus}$ is
about $10^{16}$, $(N_{\gamma}^{(0)}/ N_\nu^{(0)})$ is much smaller than
$10^{-20}$ from   Eq.$(\ref{diff-rate6})$. The density is assumed same
for both detectors, and the gamma detector is about $1$ Kg,
whereas the neutrino detector is more than $10^{6}$ Kg, and $ L(\gamma)/
L(\nu)$ is smaller than $10^{-6}$. Accordingly,
$N_{\gamma}(E_{\gamma},t)/ N_\nu(E_{\nu},t) \ll 10^{16-20-6}=10^{-10}$
  Thus the prompt gamma is not detectable.

These gamma rays actually interact  strongly with the electrons moving
parallel, which were  produced in the Supernovae, through  
the Compton or  the Thomson processes, as in
 Fig.1. These  have the enhanced probability and the photon
loses the substantial energy. These overlapping photon and electron move 
with a central velocity $ {\vec v}_0={\sigma_{e} v_{\gamma}+
\sigma_{\gamma} v_e \over \sigma_e+\sigma_{\gamma}}$ , following the classical trajectory
condition \cite{Ishikawa-Shimomura,Ishikawa-tobita-II}. The  wave packet 
size of the electron, which is mainly the thermal one, is either
macroscopic of the the size of the Supernovae or microscopic. In both cases,  $ {\vec v}_0$ is much less
than the velocity of the light, because the electron's velocity is 
substantially lower than the speed of the light, as  $v/c=10^{-3}$ or 
$v/c=10^{-1}$ for the energy KeV or $100$ KeV. Thus the velocity $ {\vec v}_0$ is
 much lower than the speed of the light.
 Consequently the signal delays by a huge amount of
 time compared with the free photon of lower energy 
different from Eq.$(\ref{diff-rate6})$.   
  The emergence of the low energy delayed  photons  instead of
  the prompt gamma rays is in accord with the observations. The detailed
  study
  of this process is outside of the scope of the present paper  and will be studied elsewhere. 
\begin{figure}[t]
\includegraphics[scale=0.35]{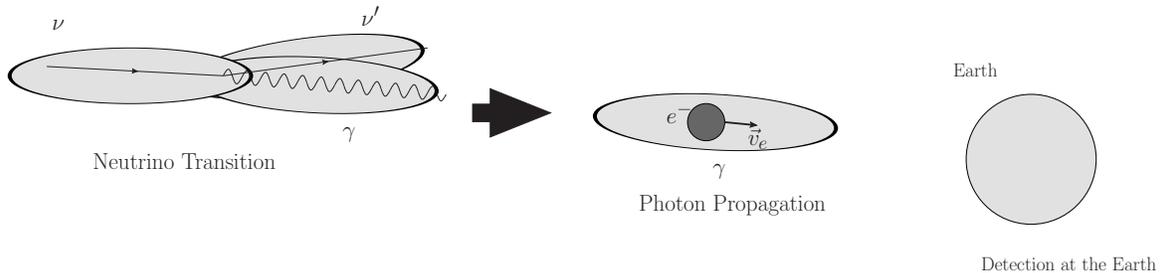}
\caption{The  photon from the neutrino decay has a large size and
 interacts with electrons by $P^{(d)}$ in the forward direction.}
\label{fig:photon-propagation}
\end{figure}
\subsection{Expanding supernovae}
 
From Eq.$(\ref{survivable })$, $P^{(d)}$ depends on the radius $R$,
and  is  negligibly small  in the region $R \ll R_0$. The SN1987A
neutrinos  reaches  the Earth unaffected by the
Galaxy. 
The neutrino flux at the ground detector agrees with that of the
initial neutrino. 
At a larger $R$, the effect becomes
prominent, and in $R \approx R_0$ or
$R \geq R_0$  $P^{(d)}$ becomes sizable, and the neutrino flux in the Earth
decreases.  Because $R_0$ is proportional to
$E_{\nu}^{-3/2}$, the reduction rate increases in the high energy. The energy
spectrum at $R > R_0$ becomes soft. 

If  the radius expands  in time  with a speed $v_{sv}$ and an initial radius
$a_0$,
\begin{eqnarray}
R=a_0 +v_{sv}t, 
\end{eqnarray} 
the survival probability varies with time.
Their magnitudes are considered as
\begin{eqnarray}
& &a_0 =10 \text{Km}, \\
& &v_{sv}=4.5 \times 10^6-3\times 10^7
 \text{meters}/\text{second}, \label{shock-velocity}
\end{eqnarray}
for the shock wave model.  The velocity is around $1/10$ of 
the light velocity  for the shock wave, and the maximum value allowed
from the causality is the
light velocity.  $a_0$ is considered small generaly.

Now we compare the theory with the observations.  Parameters in the 
theoretical expressions  Eq.$(\ref{survivable })$ are the filling
fraction and the size $\sigma_{\nu}$. Those in the
Galaxy are known but those in LMC are unknown. So we compare the
theoretical value from the Galaxy with the data \cite{Kamioka,IMB,Baksan}. 
 
Numerical simulations of supernovae explosion show that the total 
neutrino flux decreases rapidly with time but the energy spectrum 
remains or becomes wider  
at $t \leq 15$ seconds \cite{Kotake,Mayle,Totani}. The 
average neutrino energy is either constant in or slightly increasing  
with time.     Here focus to the average neutrino energy, and compare
the theory with the observations. We study the simplest case  that the SN1987A
$\phi(E_{\nu};AN1987A)$ does not vary with time \cite{Pagliaroli}. The
SN neutrino flux receives the absorption in the Galaxy and the flux
detected by the ground detector is  
\begin{eqnarray}
\Phi_{ground}^{(d)}(E_{\nu},t)= \phi(E_{\nu};AN1987A) \times { 1 \over 1+({a+
 v_{s} t
 \over R_0})^2} \label{ energy-time-spectrum}.
\end{eqnarray}
Due to low statistics, we compare a variation of 
the average energy in the period $2 \leq t \leq 12$ second.   The
average neutrino energy at $R \gg R_0$ from Eq.$(\ref{
energy-time-spectrum})$ or from Eq.$(\ref{ energy-time-spectrum-n})$ is 
\begin{eqnarray}
\langle E_{\nu} \rangle =16.7 MeV,
\end{eqnarray}
\begin{eqnarray}
\langle E_{\nu} \rangle =30 MeV.
\end{eqnarray}
They are compared with the observations in Fig.(2 ).
\begin{figure}[t]
 \includegraphics[scale=.4,angle=-90]{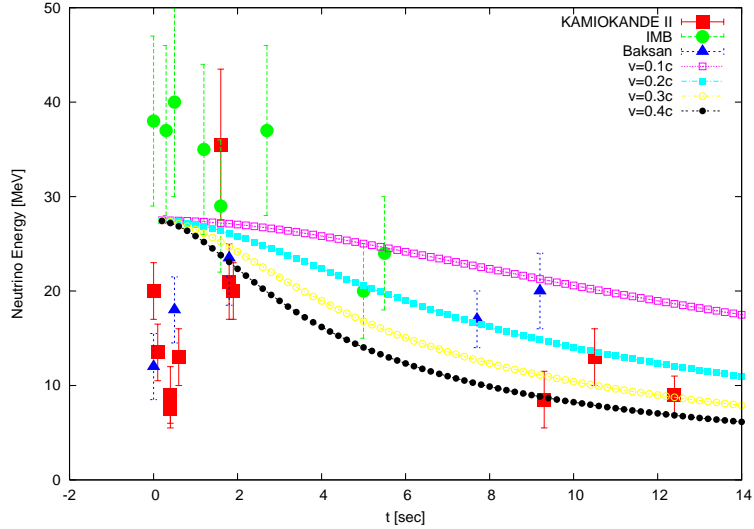}
\caption{Time dependence of neutrino energy form SN1987A is compared with those computed from Eq.$(\ref{ energy-time-spectrum})$.}
\label{fig:mu-decay-at-rest}
\end{figure}
 Our theory is in 
agreement with  the  observations, if the velocity is about one tenth of
the light velocity. Data seems to show a reduction of the higher 
energy neutrino.  From the fit, we have
 \begin{eqnarray}
v_s= 6 \times 10^7 \text {Meters}/\text{Seconds}(=0.2 c)  \pm  \delta v,\label{s-velocity}
\end{eqnarray}
which is slightly larger than  the theoretical shock front velocity
Eq.$(\ref{shock-velocity})$. $\phi(E_{\nu};AN1987A)$ may change
differently from Eq.$(\ref{ energy-time-spectrum-n})$ within
ten seconds, then the velocity  Eq. $(\ref{s-velocity})$ should be
considered the upper 
bound.  

The neutrino spectrum 
 \begin{eqnarray}
\Phi_{ground}^{nor}(E_{\nu},t)= \phi(E_{\nu};SN1987A) \times e^{-{t
 \over \tau}},  \label{ energy-time-spectrum-n}
\end{eqnarray}
 is normaly considered, where $\tau$ is the relaxation time, and 
does not show the time-dependent average energy, which does not agree
with the observation.

The energy transfered from the neutrino to the gamma ray   is stored in
extremely large waves of matters of the size $\pi R^2$. This photon
interacts with another matter wave and is not directly
detected. Assuming  that is detected directly by  the detector composed
of bound atoms, we estimate the probability of the  event. That  is  
proportional to these sizes, and is too small to detect from
Eqs.$(\ref{diff-rate6})~\text{and}(\ref{gamma-detection})$. Accordingly 
the present theory is 
consistent with the non-observation of the gamma rays from SN1987A \cite{GRS}.   

{\bf Table of physical quantities in the SN1987A and  the galaxy}

1.the density of neutral atoms; $n_{neutral}\approx 1[\text{m}^{-3}$]

2.the density of electrons; $n_e= 10^4 [\text{m}^{-3}$]

3.the magnetic field; B [Tesla]=$10^{-10}-10^{-9}$ [Tesla] \cite{galaxy-magnetic-field}

4.filling ffraction; $0.4-~0.04 /\text{m}$

5. typical radius; $R_0,\ 10^{7}$ m
\subsection{Other processes}
The enhanced $P^{(d)}$ is studied further in systems other than 
the SN1987A neutrino.
\subsubsection{Dilute gas}
High energy neutrinos produced inside the galaxy emit the photon by 
the electroweak Hall interaction  with the probability 
$P^{(d)}$ and lose their energies. The energy carried
originally by the neutrino is partly transmitted to the photon first and
to electrons, molecules, or larger objects later in the galaxy.
 These photons are out of equilibrium,
and do not follow the Planck distribution.  
At a higher energy, $P^{(d)}$ becomes larger. The high energy neutrino
has a large component of $\nu+\gamma$ in the galaxy or in the earth's
ionosphere. These would  be observed by large ground detectors such as
Icecube, Telsescope Array, and   . Owing to the geometry
dependence and other unique features of $P^{(d)}$, careful analysis 
is required.

\subsubsection{Dense gas}
In a star of  high density, 
the photon's effective mass is  larger and satisfies 
\begin{eqnarray}
m_{\gamma} > m_{\nu}+m_{\nu'},
\end{eqnarray}
hence the transitions 
\begin{eqnarray}
\nu+\bar \nu' \rightarrow \gamma, ~\gamma \rightarrow \nu+\bar \nu' \label{new-transition}
\end{eqnarray}
occur.  $\Gamma$ of these processes from the  anomalous moments 
have been studied, and a  weaker constraint than those of
\cite{GRS,kolb-turner,Bludman} 
was obtained  \cite{Fukugita-Yazaki}.  
If this  star has the magnetic field,  the electroweak Hall effect takes 
place.  However, $\sigma_{\gamma}$ in the high density is much smaller 
compared with that in the galaxy, $P^{(d)}$ is not much enhanced.  
\section{Summary}
The anomalous radiative transition of the anti-neutrino from the
supernovae 1987A in the Galaxy is studied. The survivable 
probability observed in the earth reflects the transition and 
causes the distortion of the  energy spectrum.  The theoretical 
mean energy agrees with the observations if the expanding velocity of 
the  region 
that the neutrino burst takes place is $1/5 c$.  This velocity is 
slightly larger than the standard shock front velocity, $v =0.01-0.1 c$.  
There are two possibilities to reconcile the disagreement.  One is to
 include the absorption in the LMC, and the other is to modify the shock
 front propagation. In the former one, the filling fraction in LMC,
 which is unknown now, can be estimated. Theoretical results become to
 agree with
 the observations, if  they are larger than the
 values in the Galaxy. The best fit is obtained with 
\begin{align}
\nu^{(4)}(\text{LMC})= 10\times \nu^{(4)}(\text{Galaxy}).
\end{align} 
This value is understandable from the size of LMC and and the period of
 its rotation.
 LMC has the size of one third and the    period of three times 
of those of the Galaxy. Then the
filling fraction is expected to be about ten times of the Galaxy, which
seems consistent. The second possibility of the higher speed of 
the shock front suggests that the dynamics of the shock front is
modified. This may in fact happen if $P^{(d)}$ is included  
in the shock front dynamics.

The strength of the electroweak Hall effect is determined by the filling
fraction $\nu^{(4)}$ and sizable in the system of 
the low electron density and weak magnetic field, if  their ratio 
is sizable.   The  anomalous 
transition  $\nu \rightarrow \nu'+\gamma$ of  enormously
 enhanced probability is induced, and  gives the sizable effect
 to  the neutrino from SN1987A.  
The density and the magnetic field are extremely low but their ratio is
not so small in fact in the Galaxy.   The 
transition  probability from this interaction is, $P=P^{(d)}$, instead 
of the standard $P=\Gamma T$, and    
does not increase with time interval, hence the present analysis 
differs drastically from the previous one.  $P^{(d)}$ is
proportional to the overlap of wave functions, which extends to the
gigantic area, and enhanced anomalously. The theoretical energy 
spectrum in the time interval $T \leq 12 ~\text{Seconds}$ varying 
with time rapidly is consistent with the previous experiments  
and observations, and gives the unique information through the  
survivable probability $P(\nu \rightarrow \nu)$ on  the SN1987A
radius. 
The expanding speed of the exploding star, obtained from the comparison of our
theory with the observations,  is in agreement with the speed of the 
shock front. 

The  small detection  probability of the prompt gamma from   the
process $\nu \rightarrow \nu'+\gamma$ is  in accord with the
non-observation of the prompt gamma rays during the neutrino
burst by the Solar Maximum Mission (SMN) Gamma Ray Spectrometer (GRS)
\cite{GRS,kolb-turner,Bludman}. These photons interact with matters in
the Galaxy and produce the  delayed   gamma-rays, x-rays, and others.  
Those that are produced by  the interaction of the high energy
gamma with matters in the Galaxy through $P^{(d)}$ will be studied 
in a subsequent publication.

\begin{acknowledgments}
This work was partially supported by a Grant-in-Aid for Scientific Research (Grant No. 24340043).
Authors thank Dr. Kobayashi, Dr. Nakaya, Dr. Nishikawa, and Dr. Maruyama for useful discussions on the near-detector of T2K
 experiment, Dr. Asai, Dr. Kobayashi, Dr. Mori, and Dr. Yamada for useful discussions on interferences,
 Dr. Kinoshita and Dr. Nio for useful discussions on muon g--2 experiment.
\end{acknowledgments}
\appendix

\section{ Integrals }
The light-cone singularities 
used   are partly given in many textbooks and in
Ref. \cite{Ishikawa-Tobita-PTEP,Ishikawa-Tobita-ANA}.

The integral over the coordinates $x_1$, $x_2$ and $\vec{X}_{\nu_e}$ 
is  written as 
\begin{align}
& \int d\vec{X}_{\nu}\int d^4x_1d^4x_2e^{ip_{\nu}\cdot\delta x} f(\delta x)\prod_{i=1,2}
w(x_i,X_\nu;\sigma_\nu)w(x_i,X_{\nu'};\sigma_{\nu'})
\nonumber\\
&=\left(\frac{\pi\sigma_\nu\sigma_{\nu'}}{\sigma_\nu+\sigma_{\nu'}}\right)^\frac{3}{2}
\int d\vec{X}_{\nu}e^{-\frac{\left(\tilde{\vec{X}}_\nu-\tilde{\vec{X}}_{\nu}\right)_T^2}{\sigma_\nu+\sigma_{\nu'}}}
 \int dt_1dt_2d\delta\vec{x}\,e^{ip_{\nu'}\cdot\delta x}
e^{-\frac{1}{4\sigma_\nu}\left(\delta \vec{x}-\vec{v}_\nu\delta t\right)^2
-\frac{1}{4\sigma_{\nu'}}\left(\delta\vec{x}-\vec{v}_{\nu'}\delta t\right)^2}\nonumber\\
&\times
\exp\left[-\frac{\left(\vec{v}_\nu-\vec{v}_{\nu'}\right)^2}
{\sigma_\nu+\sigma_{\nu'}}
\left(
\frac{t_1+t_2}{2}-\tilde{T}_L
\right)^2 \right]f(\delta x),\label{arb-spacetime-int}\\
&\tilde{T}_L = 
\frac{\left(\vec{v}_\nu-\vec{v}_{\nu'}\right)
\cdot\left(\tilde{\vec{X}}_\nu-\tilde{\vec{X}}_{\nu'}\right)}
{\left(\vec{v}_{\nu}-\vec{v}_{\nu'}\right)^2},\nonumber
\end{align}
and using Gaussian approximation for integration in $\vec{X}_{\nu'}$, we have
\begin{align}
 \left({\pi^2\sigma_\nu\sigma_{\nu'}}\right)^\frac{3}{2}
 \int dt_1dt_2d\delta\vec{x}\,e^{ip_{\nu'}\cdot\delta x}
e^{-\frac{1}{4\sigma_\nu}\left(\delta \vec{x}-\vec{v}_\nu\delta t\right)^2
-\frac{1}{4\sigma_{\nu'}}\left(\delta\vec{x}-\vec{v}_{\nu'}\delta t\right)^2}
f(\delta x).\label{app-Xint}
\end{align}
For $f(x) = i\frac{\epsilon(\delta t)}{4\pi}\delta(\lambda)$, integral in Eq. \eqref{app-Xint} is written as
\begin{align}
& \int dt_1dt_2 d\delta\vec{x} e^{ip_{\nu'}\cdot\delta
 x}e^{-\frac{1}{4\sigma_
\nu}\left(\delta\vec{x}-\vec{v}_\nu\delta t\right)^2-\frac{1}{4\sigma_{\nu'}}
\left(\delta\vec{x}-\vec{v}_{\nu'}\delta t\right)^2
}\frac{i}{4\pi}\delta(\lambda)\epsilon(\delta t)
\nonumber\\
&= \int^T_0 dt_1dt_2\,e^{-\frac{(\vec{v}_\nu-\vec{v}_{\nu'})^2\delta t^2}{4\sigma_\nu}}\int d\vec{r}e^{ip_{\nu'}\cdot\delta x}
e^{-\frac{\left(\vec{r}-\vec{v}_{\nu'}\delta t\right)^2}{4\sigma_{\nu'}}}
\frac{i}{4\pi}\delta(\lambda)\epsilon(\delta t)
\nonumber\\
&\simeq \frac{i}{2}\sigma_{\nu'}\int_0^Tdt_1dt_2e^{-\frac{(\vec{v}_\nu-\vec{v}_{\nu'})^2}{4\sigma_\nu}\delta t^2}
e^{-\frac{(1-|\vec{v}_{\nu'}|)^2}{4\sigma_{\nu'}}\delta t^2}\frac{e^{i\omega_{\nu'}\delta t}}{\delta t},\label{app-Xint-delta}
\end{align}
where $\omega_{\nu'} = \frac{m_{\nu'}^2}{2E_{\nu'}}$, and $\sigma_{\nu'}|\vec{p}_{\nu'}|\ll T$ is used.
Due to the small mass of neutrino, $e^{-\frac{(\vec{v}_\nu-\vec{v}_{\nu'})^2}{4\sigma_\nu}\delta t^2}=e^{-\frac{(1-|\vec{v}_{\nu'}|)^2\delta t^2}{4\sigma_{\nu'}}} = 1$, but this suppression
factor cannot be ignored for massive particles.

\section{Universal function $\tilde{g}(\omega,T)$ }\label{app-gtilde}
Due to the small mass of neutrino, the approximation
$e^{-\frac{(\vec{v}_\nu-\vec{v}_{\nu'})^2}{4\sigma_\nu}\delta
t^2}=e^{-\frac{(1-|\vec{v}_{\nu'}|)^2\delta t^2}{4\sigma_{\nu'}}} = 1$
is good,
which  cannot be used for massive particles, and we have
\begin{align}
 g(\omega_{\nu},T) 
&={i}\int_0^Tdt_1dt_2\frac{e^{i\omega_\nu(t_1-t_2)}}{t_1-t_2}
\nonumber\\
&=-2 (\int_0^T dt\frac{\sin(\omega_\nu t)}{t}-\frac{1-\cos(\omega_{\gamma}T)}{\omega_{\gamma}T})
,\label{lifetime-g}
\end{align}
where $t_+=\frac{t_1+t_2}{2},\ t_- = t_1-t_2$. 
Since $g(\omega_{\nu},\infty)=-{ \pi }$ is cancelled with the short-range term 
from $J_\text{regular}$,
we write  
\begin{align}
 \tilde{g}(\omega_\nu,T) = { \pi }-2(
\int_0^T dt\frac{\sin(\omega_\nu t)}{t} -\frac{1-\cos(\omega_{\gamma}T)}{\omega_{\gamma}T}).\label{lifetime-gtilde}
\end{align}
\begin{align}
& \tilde{g}(\omega,T)| \sim \frac{2}{\omega T}; ~~~\omega T\gg 1.\label{asymptotic-gtilde1}
\end{align}

\section{integral}
The integral over the relative coordinates  is given by
\begin{eqnarray}
 & &\int d{\vec r} e^{ip{\vec n_2}{\vec r}} e^{-{ 1 \over 2\sigma}({\vec
  r}-{\vec v}_1 t)^2} \delta(r^2-c^2t^2)\\
 & &=\int d{\vec s}e^{ip{\vec n_2}({\vec v}_1 t + {\vec s})} e^{-{ 1 \over 2\sigma}({\vec
  s})^2} \delta(s^2+v_1^2 t^2 -c^2t^2+2{\vec s}{\vec v}_1 t)\nonumber \\
 & &=e^{ip{\vec n_2}({\vec v}_1 t)} \int d{\vec s} e^{ip(-{\vec
  v}_1+{\vec v}_2) {\vec
  s}} e^{ip({\vec v}_1{\vec s}) -{ 1 \over 2\sigma}({\vec
  s})^2}  \delta(s^2 +2{\vec s}{\vec v}_1 t) \nonumber \\
& &=e^{ip{\vec n_2}({\vec v}_1 t)} \int d{\vec s} \sum_l {1 \over l!}(ip(-{\vec
  v}_1+{\vec n}_2) {\vec
  s})^l e^{ip({\vec v}_1{\vec s}) -{ 1 \over 2\sigma}({\vec
  s})^2}  \delta(s^2 +2{\vec s}{\vec v}_1 t) \nonumber \\
 & &=e^{ip{\vec n_2}({\vec v}_1 t)} \int 2\pi d \cos \theta s^2 d s
  e^{ip v_1 s \cos \theta } e^{-{ 1 \over 2\sigma}
  s^2} ( \delta(s^2+2sv_1 t \cos \theta) (1 +\epsilon) \nonumber \\
& &=e^{ip{\vec n_2}({\vec v}_1 t)} \int 2\pi {1 \over 2s |t| v_1} s^2 d s
  e^{-ip v_1 s { s^2 \over 2stv_1}} e^{-{ 1 \over 2\sigma}
  s^2}  (1 +\epsilon) \nonumber \\
 & &=e^{ip{\vec n_2}({\vec v}_1 t)} 2\pi {1 \over 4 |t| v_1} \int d s^2
  e^{-({1 \over 2\sigma}+{ip  \over 2 t })s^2}(1 +\epsilon)\nonumber \\
 & &= e^{ip({\vec n_2}\cdot {\vec v}_1) t} 2\pi {1 \over 4 |t| v_1} {1
  \over {1 \over 2 \sigma}+{ip \over 2t }}(1 +\epsilon), \nonumber
\end{eqnarray}
where the variable $\vec s={\vec r}-{\vec v}_1 t$ is used and 
 a small quantity $\epsilon$
is ignored.

The integral over the times 
\begin{eqnarray}
\int_0^T dt_1 dt_2 e^{-i(E  -p({\vec n{_2}{\vec v}_1})) t} {1 \over 4 |t|
 v_1} {1
  \over {1 \over 2 \sigma}+{ip \over 2t }}(1 +\epsilon) sign ~t, \nonumber
\end{eqnarray}
in the region ${p \over t} \ll {1 \over 2\sigma}$ is,
\begin{eqnarray}
\int_0^T dt_1 dt_2 e^{-i(E  -p({\vec n{_2}{\vec v}_1})) t} {1 \over 4 t
 v_1} {1
  \over {1 \over 2 \sigma}}=-i {\sigma \over 2v_1} T (\pi/2-{\tilde g(\tilde
  \omega T)})\nonumber
\end{eqnarray}
and in the region ${1 \over 2\sigma} \ll {p \over t} $ is,
\begin{eqnarray}
\int_0^T dt_1 dt_2 e^{-i(E  -p({\vec n{_2}{\vec v}_1})) t} {1 \over 4 t
 v_1} {1
  \over {ip \over 2t }}= { -i \over  2p v_2} 4 {(\sin \tilde \omega T)^2
  \over \tilde \omega^2}, \nonumber
\end{eqnarray}
where $\theta$ is the angle between  ${\vec n}_2$ and ${\vec v}_1$ and
\begin{eqnarray}
\tilde \omega=E(p)-p{\vec n}_2{\vec v}_1=p(1 - \cos \theta)+{m^2  \over 2E}.
\end{eqnarray}

\end{document}